\documentclass[conference]{IEEEtran}
\IEEEoverridecommandlockouts
\usepackage{cite}
\usepackage{amsmath,amssymb,amsfonts}
\usepackage{algorithmic}
\usepackage{graphicx}
\usepackage{textcomp}
\usepackage{xcolor}
\def\BibTeX{{\rm B\kern-.05em{\sc i\kern-.025em b}\kern-.08em
    T\kern-.1667em\lower.7ex\hbox{E}\kern-.125emX}}
\begin{document}

\title{Robust Machine Learning in Critical Care –- Software Engineering and Medical Perspectives}

\author{\IEEEauthorblockN{Miroslaw Staron}
\IEEEauthorblockA{\textit{Computer Science and Engineering} \\
\textit{Chalmers $|$ University of Gothenburg}\\
Gothenburg, Sweden \\
miroslaw.staron@gu.se}
\and
\IEEEauthorblockN{Helena Odenstedt Herg{\`e}s, Silvana Naredi, \\ Linda Block, Ali El-Merhi, Richard Vithal}
\IEEEauthorblockA{\textit{Dep. of Anesthesiology and Intensive Care} \\ \textit{Sahlgrenska Academy} and \\
\textit{Region Västra Götaland} \\ \textit{Sahlgrenska University Hospital} \\ \textit{Dep. of Anesthesia and Intensive Care}, \\
Gothenburg, Sweden \\
(name.surname)@vgregion.se}
\and 
\IEEEauthorblockN{Mikael Elam}
\IEEEauthorblockA{\textit{Institute of Neurosc. and Physiology} \\ \textit{Dep. of Clinical Neuroscience} \\
\textit{Sahlgrenska Academy}, and \\
\textit{Dep. of Clinical Neurophysiology} \\
\textit{Sahlgrenska Univ. Hospital}, and \\
\textit{MedTech West} \\
\textit{Sahlgrenska Univ. Hospital} \\
Gothenburg, Sweden \\
mikael.elam@gu.se}
}

\maketitle

\begin{abstract}
Using machine learning in clinical practice poses hard requirements on explainability, reliability, replicability and robustness of these systems. Therefore, developing reliable software for monitoring critically ill patients requires close collaboration between physicians and software engineers. However, these two different disciplines need to find own research perspectives in order to contribute to both the medical and the software engineering domain. In this paper, we address the problem of how to establish a collaboration where software engineering and medicine meets to design robust machine learning systems to be used in patient care. We describe how we designed software systems for monitoring patients under carotid endarterectomy, in particular focusing on the process of knowledge building in the research team. Our results show what to consider when setting up such a collaboration, how it develops over time and what kind of systems can be constructed based on it. We conclude that the main challenge is to find a good research team, where different competences are committed to a common goal.  
\end{abstract}


\section{Introduction}
\noindent
Designing reliable software systems for monitoring critically ill patients is a complex process, which often requires in-depth understanding of the clinical practice, data science, software engineering and involves strict certification processes \cite{forsstrom1997certification}. The requirements on software systems include repeatability and explainability of results, traceability of the recommendations and robustness to changes in data and scalability of the software \cite{arpteg2018software}.  These requirements need to be addressed already from the first steps of the design, as the transparency of data collection, labelling of data points and the mitigation of confounding factors cannot be added in late software development phases. Software design processes need to take into  consideration both the state-of-the-art practices in software engineering and the medical practices that affect the use of the software. 

The research behind the clinical event identification, designing data analysis pipelines and the subsequent machine learning models require close collaborations between the software engineering and the physicians. Therefore, efficient research teams in this area are often cross-disciplinary, which is the best-case scenario, but also the scenario that is the most difficult to achieve \cite{maslove2021}. One of the impediments in the establishment of such teams is the ability to find common goals, research questions and publication venues. For the researchers in the medical domain, the focus is often to find new methods to improve medical care, or to develop more efficient and reliable treatments. For the researchers in software engineering, the focus is instead on finding methods for improving software engineering practices, such as requirements engineering, modelling, testing or quality assurance. 

In our study, we set off to find evidence on what makes the collaboration between software engineers and physicians effective when designing machine learning systems for use in critical care. In particular, we studied how to establish such effective and efficient research teams by addressing the following research question: \emph{What has to be taken into consideration when establishing an efficient cross-disciplinary collaboration between software engineers and physicians?}

We used chronological analysis of our research journal, which we kept during the studies. The goal of these studies is to improve the detection of cerebral ischemia using artificial intelligence (AI)  \cite{block2020cerebral}. The results from that analysis  show that the close, on-site collaboration, frequent presentation of results and using explainable machine learning algorithms were keys to success.  

The remaining of the paper is structured as follows. Section \ref{sec:related_work} presents the most important related work. Section \ref{sec:research_context} describes our collaboration, the research project, its goals and how we collected the data for our retrospective analysis. Section \ref{sec:phases_of_collaboration} presents the results organized by the phases of the collaboration. The results are in form of the knowledge development and the research topic development over time. We summarize the recommendations for other teams in Section \ref{sec:recommendations}. Finally, Section \ref{sec:conclusions} summarizes the paper and presents conclusions. 

\section{Related work}
\label{sec:related_work}
\noindent
In the field of software engineering, the studies on collaborative projects are most often focused on the collaboration between software engineering researchers and practitioners. An example of such a study is the work of Sandberg et al. \cite{sandberg2011agile}, who studied the collaboration between a group of researchers and one company, which was later extended to multiple companies \cite{sandberg2017meeting}. Similar studies were reported in specific areas, such as testing \cite{garousi2017industry} or design, modelling and project management \cite{garousi2019characterizing}. This line of research is focused on one domain and the ability to align software development goals with the software engineering research goals. In some sense, however, these goals are already close as both parties want to improve the software engineering practices. Examples of resulting factors are engagement, co-production of results and goal alignment. Our work provides a different perspective on this problem -- collaboration between researchers from different domains where the goals cannot be as easily aligned as within one domain, or even contradictory at times. 

In the same domain, one line of research is focused on the transfer of research results, i.e., where innovation is created by researchers and adopted by the practitioners. An example of such research area is the work of Zahedi et al. \cite{zahedi2018empirical}, Gorschek et al. \cite{gorschek2006model} and Mikkonen et al. \cite{mikkonen2018continuous}. The main contributions in this area help to increase the level of utilization of research results. We contribute to this line of research by providing an account of how to increase the utilization of machine learning models in clinical practice. 

Another line of research is focused on the collaboration between academia (university) and industry (companies). In this line of research the focus is on the alignment of goals and co-production of research results, and on the involvement of students and their learning. An example of this kind of research is the work of Awasthy et al. \cite{awasthy2020framework} and Ross and Riley \cite{ross2018degree}. Although there is a significant progress in this area, in particular when discussing scalability of the results, there are still unanswered questions on how to build common awareness of the problem or how to transfer technology between domains (software engineering specialists vs. clinical specialists). Our work is focused on this kind of technology transfer. 

A typical way of collaborating and addressing multidisciplinary questions is by using observations, e.g., non-participatory observations using video recordings. Ivarsson and {\AA}berg \cite{ivarsson2020role} provide a recent example of such a study, where the information systems researchers studied communication challenges during surgery. These studies provide valuable examples of how to increase understanding of clinical practices, but are often focused on one domain only. Our work aims at addressing both the engineering and the medical domains simultaneously, in particular in finding relevant research questions for both domains.

Finally, the interest for using machine learning in critical care units has been growing rapidly in the last few years. The main focus so far is on the utilization of the patient data \cite{poncette2020improvements}, ethical aspects \cite{beil2019ethical} and managing of high frequency patient data \cite{goodwin2020practical}. All of these advancements require collaboration between software engineers/data science specialists and critical care physicians, as recognized by Maslove et al. \cite{maslove2021}. Our work contributes with practical guidelines and recommendations on how to set up such collaboration and how to develop common knowledge for all participating domains. 

\section{Research context}
\label{sec:research_context}
\noindent
To understand the challenges in establishing the collaboration, we decided to collect data from our research project on a daily basis. We used MS OneNote tool to record important events and to reflect on them. We used the project plan as a starting point for the journal, and we used a similar approach as previous research studies \cite{mathiassen2013professionally}.

\begin{figure}[!htb]
    \centering
    \includegraphics[width=\columnwidth]{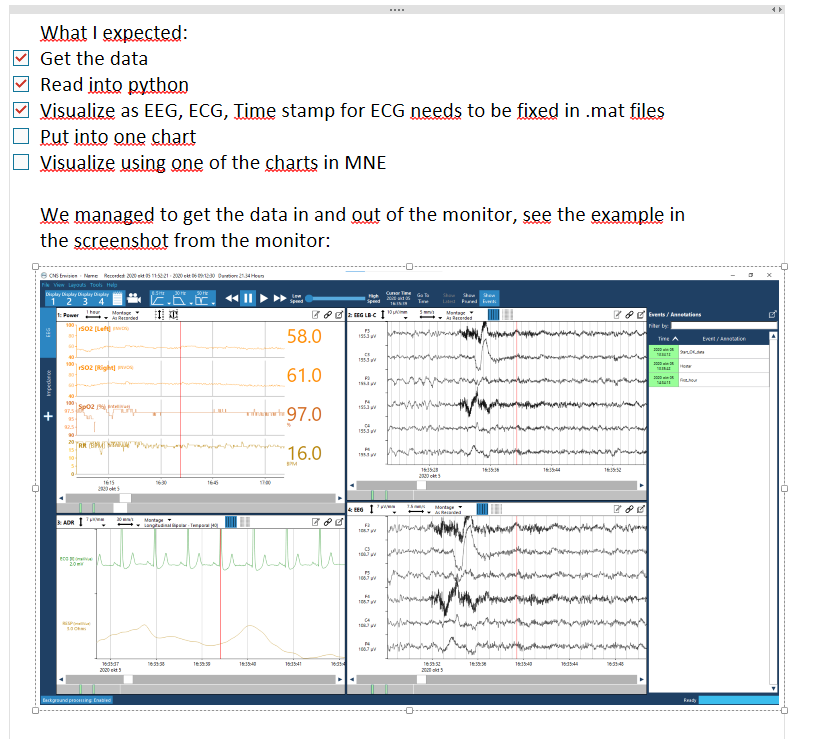}
    \caption{Example of a part of journal entry for one day, screenshot from MS OneNote. The entire entry is approx. 2 pages long. Expectations and notes are above the diagram. The diagram is a screenshot of the data collection program connected to the Moberg monitor.}
    \label{fig:my_label}
\end{figure}

Everyday, we started by setting up expectations for the day, which we set during a joint team meeting. Then, during the day, we focused on the activities planned for that day. At the commence of the day, we recorded all observations which we deemed relevant, and shared with the research team. An example of a journal entry is presented in Figure \ref{fig:my_label}.

The format of the journal was very open ended, with the only required part being the expectations. The free form of OneNote was perfect to include pictures taken during clinical visits, screenshots of the analysis tools or illustrations from the textbooks used in the learning phase. 

Table \ref{tab:summary} presents the expectations and the main findings for selected days throughout the project. 

\begin{table*}[!htb]
    \footnotesize
    \centering
    \begin{tabular}{||p{0.5cm}|p{7cm}|p{10cm}||}
        \hline
        \textbf{Day} & 
        \textbf{Expectations} & 
        \textbf{Main Findings} \\ \hline
        
        1 & 
        Learn how to use equipment (Moberg monitor) and learn how to read EEG (Electroencephalograph) signals. &  
        1) Data analysis pipeline requires some manual intervention (export from the Moberg monitor), and 2) EEG signals require more expert knowledge that initially expected. \\ \hline
        
        2 & 
        Visit the ICU (Intensive Care Unit) and get familiar with the equipment used. & 
        1) Heart Rate Variability patterns may differ significantly, 2) ECG signals can differ in the amplitude based on which lead is displayed.\\ \hline
        
        3 & 
        Export data from a pilot-patient and check how we can create a data analysis pipeline. &  
        1) Using the data in Python works best with the HeartPy toolkit, 2) Clinical events need to be scrutinized and taken into consideration when evaluating data to increase quality. \\ \hline
        
        4 & 
        Add clinical events to the data and create the feature vector for the ECG signal. &  
        1) ECG is the most robust signal and the EEG signal is the one that is most prone to disturbances (even moving a cloth on patient's head can cause artefacts), 2) qEEG (quantitative EEG) is quite straightforward using the standard Python \texttt{fft} function. \\ \hline
        
        5 & 
        Visualize the ECG data using HeartPy package and visualize the features extracted from ECG. &  
        We need to set-up the HeartPy analysis almost individually to avoid confusing R- and T-peaks in the ECG signal. \\ \hline
        
        6 & 
        Learn about the NIRS\footnote{Near-infrared Spectroscopy, \cite{villringer1993near}.} signal and extract features from the EEG signal. &  
        NIRS seemed to be a signal that is not very reliable due to the limitation of how deep into the brain tissue the measurement reaches. \\ \hline
        
        7 & 
        Collect data from the first study-patient, make the first visualization using t-SNE\footnote{t-distributed Stochastic Network Embedding \cite{maaten2008visualizing}.} &  
        1) Creating the data frame with all features resulted in 48 features. 2) After data cleaning and labelling, each patient resulted in over 100 data points. \\ \hline
        
        8 & 
        Make the first data analysis using machine learning. &  
        The accuracy of the first patient was 0.61, when recognizing all events. \\ \hline
        
        9 & 
        Compare different machine learning algorithms w.r.t. accuracy. &
        1) After testing CART Decision Trees, AdaBoost, Random Forest and Support Vector Machines, we found that the accuracy differences are very small (max 0.2), 2) we decided to use Random Forest as it provided us with the ability to plot the feature importance chart, 3) The feature importance chart helped the team to validate (explain) the ML results. \\ \hline
        
        10 & 
        Review the entire surgical procedure, including anesthesia, together with collected data, minute by minute, with the entire team. &  
        1) Reduced number of clinical events of relevance to five (ten as each event has a start and end), 2) NIRS was found to be the most clear signal for distinguishing the events both by physicians and by machine learning. \\ \hline
        
        11 & 
        Collect and analyze the data for two new patients. &  
        t-SNE visualizations, confusion matrices and ROC curves were found to be very good charts to communicate the results within the team. \\ \hline
        
        12 & 
        Collect and analyze the data from three more patients. Find a way of adding the labels to the data automatically. &  
        1) The most important signals for these two patients were related to heart rate variability, and 2) due to problems with EEG electrodes, the results were much lower than previously (the problems could not be fixed in the data analysis/cleaning phase). \\ \hline
        
        13 & 
        Prepare a summary of analyses and present to the entire team for discussion. &
        The features used for all patients were consistent with each other (i.e., the most important features were the same for all patients), which strengthened our results. \\ \hline
        
        14 & 
        Add data from the post-operative care for all patients. &  
        1) Adding the post-operative care as an event increased the prediction accuracy to 0.90, 2) we could use the post-operative period as a baseline. \\ \hline
        
        15 & 
        Present the summary to the team and analyze the EEG signals in detail for one patient. &  
        1) The results (e.g., feature importance charts) were consistent with how the physicians could recognize the events, and 2) in-depth studies of the EEG signals showed that it is very difficult to use that signal in practice by non-experienced analysis and physicians (recognizing the patterns requires experience). \\ \hline
        
        16 & 
        Identify new research questions based on the analyses and insights so far.& 
        1) We found the need to study how to identify rare events (e.g., when we only have 2 data points out of 200), and 2) we needed to study how the data collection can be done in a more emergent setting, for example for patients admitted for acute cerebral thrombectomy. \\ \hline
        
        17 & 
        Create detailed diagrams for all features collected to help to manually explore the data for each patient. &  
        The detailed diagrams are important for traceability of results, but their quantity (48 diagrams for as many features) make it difficult to examine all of them in detail for every patient. \\ \hline
        
        18 & 
        Analyze one more patient. &  
        1) This surgery included the establishment of a vascular shunt, which made the NIRS signal less clear, 2) qEEG features were the most important ones for the algorithm, 3) the accuracy was lower due to more noisy NIRS signal -- 0.82. \\ \hline
        
        19 & 
        Prepare dissemination plan and read upon the details on the surgical procedure. &  
        Planning of publications resulted in finding three new research topics of joint interest -- potentially new surgical procedures to study and new data collection methods (autolabelling of heart rate data using NIRS signals). \\ \hline
        
        20 & 
        Prepare the draft of the first presentation. &  
        The lowest accuracy was 0.82 and the highest 0.98 for the five patients that were analyzed. Based on the comparison with the existing literature, these results were better than existing work. \\ \hline
    \end{tabular}
    \caption{Summary of expectations and main findings for selected days of the project.}
    \label{tab:summary}
\end{table*}

Naturally, the planning of the project was done beforehand, but the patients' data collection was not. The studied patients underwent cartoid endarteractomy, which is a procedure conducted within 24-48 hours of admitting the patient. 

The table illustrates the gradual progression of the focus from understanding the clinical practice to the analysis of the data and planning of new studies to increase the external validity of the results. This was done on premises of the ICU, i.e., where the clinical practice takes place, and it helped the software engineers to observe the practice. 

\section{Phases of collaboration}
\label{sec:phases_of_collaboration}
\noindent
Since our notes are chronological, we decided to structure our experiences and good practices based on the phases of the collaboration -- starting from finding the right partners to making the results generalizable to new contexts. 

\subsection{Finding the right partners}
\noindent
In our view, the most essential part of the project is to find the right partners. In general, collaborations between different fields are based on individual engagement, but in the essence it is also based on the competences. 

In the case of our project, we found the collaborations by discussing common research areas outside of the clinical contexts. The physicians mentioned a research challenge related to detection of delayed cerebral ischemia after subarachnoid hemorrhage using heart variability measurements. The software engineers were interested in understanding challenges with data quality from the clinical procedures. So, we had two perspectives on the same data:
\begin{itemize}
    \item Software engineering: data quality, data collection, confounding factors (or noise) and scalability of the software. 
    \item Medical: automatic support for identification of medical events, automated monitoring and detection of specific conditions as well as better use of the large amount of data generated in ICUs and in operating rooms (OR). 
\end{itemize}

These two perspectives are distinct, but complementary. Understanding the quality of the data is essential for building high quality machine learning based systems \cite{arpteg2018software}, but it is often treated from a computational perspective -- e.g., finding noisy data points using statistics \cite{zhu2004class}. Studying the sources of noise or bad quality are, henceforth, extremely important for software engineers \cite{gudivada2017data}.

Our recommendations for this phase are, therefore:
\begin{enumerate}
    \item Identify an existing data which can be used to start the collaboration with a low entry level for both disciplines. This helps to move from the conceptual to the practical phase quickly; the practical work is where most of the learning happens. 
    
    \item When working on the problem, ask questions about the data, underlying procedures, ways of working and the analysis as often as possible. This helps to learn about each domain and understand what is important; this learning helps to design new studies that provide higher quality data (``rookie mistakes'' can be avoided the next time). 
    \item Organize a brainstorming session (or many) listing out the most important research challenges related to the problem, which can be solved by the team. This will help to understand the potential and design a series of studies; it can be published as study protocols, e.g., \cite{block2020cerebral}.
    
    \item Prepare brief (an outline) of a research design for each study. This helps to understand which elements each study should contain in order to maximize the generalizability of results; it is also important to plan which competence to include in each study. 
\end{enumerate}

Following these guidelines increases the efficiency of the research team and provides a clear view on the goals. 

\subsection{Finding relevant research questions}
\noindent
One of the main challenges in finding the right collaboration is to find a set of research questions that are of interest for both disciplines. A typical counter-example is when research questions are relevant for the medical domain, but not for software engineering -- \emph{How to recognize clinical events relevant for cerebral ischemia?} 

In order to find the relevant questions and to design studies that address them, both disciplines need to be active in the initiation/design of the research project. Being involved from the beginning allows to understand which research problems can be relevant and also how to prepare the study so that both disciplines can benefit from it. 

Both disciplines need to be explicit about the types of research problems. For example, the software engineers need to describe their context and needs, e.g., posing research questions like \emph{How to automatically identify data quality issues related to OR procedures?}

Table \ref{tab:research_questions} presents examples of research questions which are relevant for both domains, stemming from our studies. 

\begin{table}[!h]
    \begin{tabular}{||p{0.7cm}|p{3.5cm}|p{3.5cm}||}
        \hline
        \textbf{RQ} & \textbf{Medical} & \textbf{SE} \\ \hline
        RQ1 & Can risk of upcoming cererbral ischemia be detected by a monitoring system? & Which algorithms are best for recognizing cerebral ischemia? \\ \hline
        RQ2 & Is it possible to detect a risk of cerebral ischemia during carotid endarterectomy? & How to ensure that the data is clean of noise?  \\ \hline
        RQ3 & Can anesthesia mask or mimic cerebral ischemia? & Which medical information needs to be included for accurate detection of cerebral ischemia using ML? \\ \hline
        RQ4 & Which monitoring components are needed to create an alarm system for detection of cerebral ischemia?  & How to reduce false-positive alarms? \\ \hline
        RQ5 & How to create a monitoring system that is sensitive enough?  & Which machine learning performance measures are the most relevant for assessing the quality of the software system?\\ \hline
    \end{tabular}
    \caption{Examples of research questions relevant for the medical and software engineering domains.}
    \label{tab:research_questions}
\end{table}

It is important that these research questions are established at the beginning of the project, so that the research team has an awareness of each other's goals \cite{sandberg2011agile}. For a software engineer, this requires exploring research in the areas that require customer or user collaboration. In our case, we focused on exploring the aspects of differences between the machine learning algorithms in practice (e.g., comparing Random Forest with Support Vector Machines), data quality (e.g., automated data cleaning vs. manual data labelling) or scalability (e.g., cross-patient model evaluation).  

We recommend the following when finding the research questions:
\begin{enumerate}
    \item Each research team member prepares a set of research questions of interest on her/his own before sharing with others. This helps to focus on each disciplines/subdisciplines and individuals goals for the study.
    \item Discuss the research questions together and plan for them in the series of studies identified in the previous phase; this shows the progression of the research problems and their solutions.
    \item Make a preliminary plan for which research questions should lead to which publications and group them accordingly. This grouping leads to more efficient planning of the dissemination of the results from the project. 
\end{enumerate}

Once the research questions are planned, the team should start with the pilots for the first study in line. 

\subsection{Pilot studies: Understanding the environment}
\noindent
The most important aim for the pilot study is to understand each other's practice and to design a full set-up of the study at hand. In our case, we decided to conduct the study on one patient in the operating room. The physicians in the team were present during the surgery and recorded the events during the surgery. These events were important as we used them to label the data for our study. 

\begin{figure}[!htb]
    \centering
    \includegraphics[width=\columnwidth]{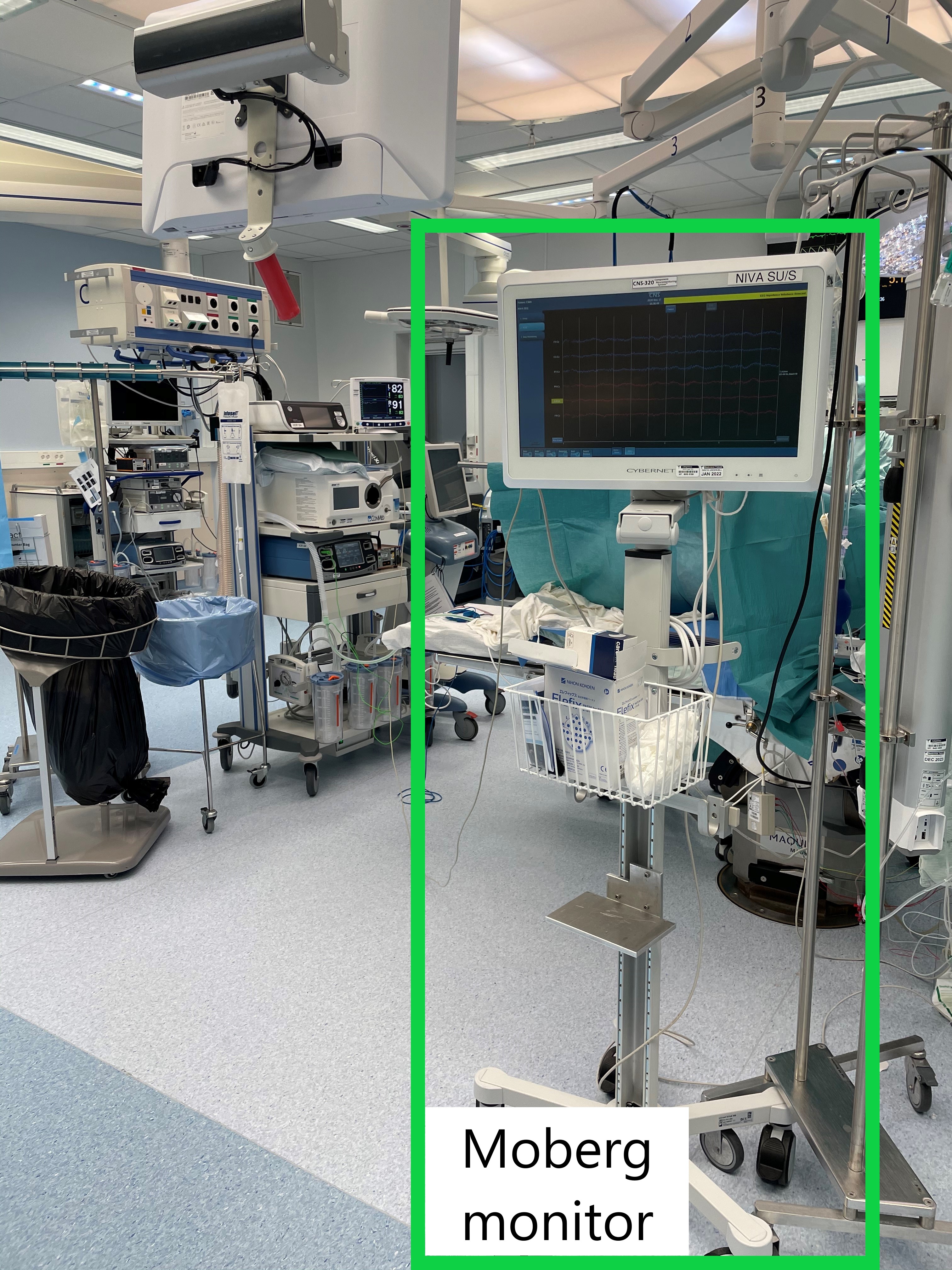}
    \caption{Clinical set-up for data collection. Moberg monitor is an additional equipment used to collect data from multiple sensors (NIRS, ECG, EEG), annotate them and store in a single file.}
    \label{fig:OR}
\end{figure}

In the pilot study, we learned about the limitations of the technology, in particular the sensors used in the study. That knowledge was important as the sensor data was needed for the analysis. The software engineers were introduced to the OR surgical procedure and the OR environment. Figure \ref{fig:OR} presents a set-up in the OR, with the Moberg monitor in the central place. 

We found that the most important part of the understanding was the concept of events during surgery. An event is the same as the label which we used for machine learning. The events show what happens during the surgery, which we want the machine learning to recognize, identify and predict. Examples of events are: 
\begin{itemize}
    \item induction of anaesthesia,
    \item increase/decrease of drugs, 
    \item clamping of carotid artery, or 
    \item post-operative care. 
\end{itemize}

The events were crucial as they provided a common ground for discussing the procedure and the resulting data. Software engineers could ask questions like: Why is the induction of anesthesia important for the data? Why should we monitor the increase/decrease of the drugs?  Does the clamping of the carotid artery influence the heart rate variability? 

Asking these questions led to the understanding of underlying confounding factors in the data. In our case, we found that the role of anesthesiologists in the OR is to increase the blood pressure to maintain cerebral perfusion, which could confound our results as we aimed to recognize differences in heart rate variability caused by closing the carotid artery. 

As a result of these discussions, in our case, we found a number of periods during the surgical procedure where certain reference periods could be established: taking six deep breaths to standardize heart rate variability, unstimulated period after anesthesia induction, additional events such as stump pressure measurement etcetera. 

From these discussions, the physicians learned what the data analysts need to see in the data to train the machine learning classifier. They have also learned about the dependencies between the signals and how the raw signals (e.g., raw EEG, Electroencephalograph) are quantified for further processing. 

\begin{figure}
    \centering
    \includegraphics[width=0.8\columnwidth]{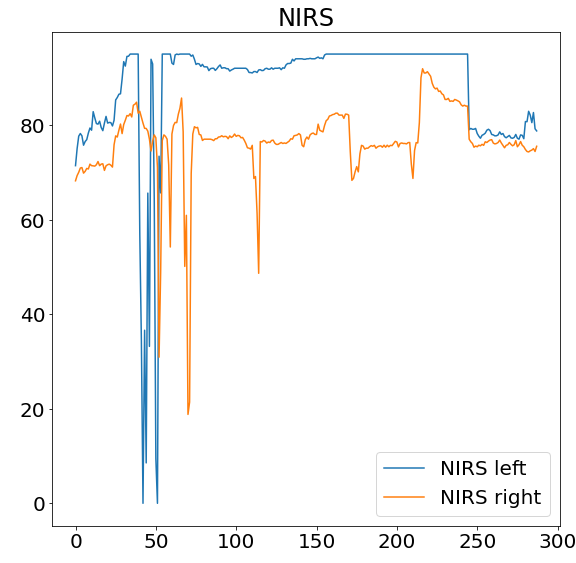}
    \caption{An example diagram -- Near-Infrared Spectography}
    \label{fig:NIRS_and_RESP}
\end{figure}

Figure \ref{fig:NIRS_and_RESP} shows an example of the result of data collection -- two signals from one patient. 

For this phase we recommend: 
\begin{enumerate}
   \item Record as many events as possible in the OR, as it is easier to remove them than to find them later on. Having more events helps the software engineers to understand what is important. 
   \item Review the entire data from the OR and select representative periods for further analysis, instead of labelling the entire data set. This helps to find the data that is clean enough for the algorithms. 
   \item Show the ``quick and dirty'' analysis at the early stages to the entire team. This helps to raise awareness of what is important for analysis, which in turn leads to better data collection. 
   \item Provide each other with the important literature on the research topic in order to introduce each other to the terminology, procedures and important research results in the field. This helps to learn the terminology, vocabulary and to find in which direction the data analysis should go (e.g., selection of algorithms, tuning of parameters). 
\end{enumerate}

After this stage, we concluded that the research team have such a level of understanding that it is virtually impossible to bring in new members without significant training effort. At this stage, both domains have gained substantial understanding of each other. 

We believe that this stage helps to build such a common understanding that it replaces the need for requirement specifications when designing prototypes. In case when the requirement specifications are to be used in the future, the software engineers have sufficient domain knowledge to be able to create quality requirement specification. 

\subsection{Data analysis}
\noindent
When analyzing the data, we found that the most important activity is the joint discussion of the results. It helps the software engineers to understand which empirical (clinical) events the data represents. It also helps the physicians to understand how the data is analyzed and therefore add important information. 

In the case of our collaboration, we presented the results on a daily basis, every time a new patient data was analyzed. This helped us to evolve the procedure -- both in the OR (identifying events, bringing in extra competence for setting up EEG electrodes), and in the analysis (finding which events are important to recognize). 

These discussions helped us to reduce the number of events monitored by half, for example: the baseline events were important, but too short to be used in the analysis (too few data points), the events of continuous drug administration that lasted the entire procedure were reduced to only two (as their effect overlapped other clinical events). 

Our analysis included a number of machine learning algorithms -- from the simple CART decision trees \cite{quinlan1986induction}, through the use of AdaBoost boosting \cite{ying2013advance}, Support Vector Machines \cite{cristianini2000introduction} to the use of Random Forest \cite{breiman2001random}. When discussing the results, we also found guidelines for reporting studies involving machine learning in medicine \cite{luo2016guidelines}, which helped the software engineers to understand the requirements for machine learning systems from the clinical practice side. 

In the data analysis phase, we recommend the following:
\begin{enumerate}
    \item Use the reporting guidelines from the beginning. Using the reporting guidelines from the medical literature helped us to understand how to process the data to minimize confounding factors related to data analysis (conclusion validity). 
    \item Optimize the data labelling process with machine learning in mind: make sure that the data can be machine processed as much as possible. This helps to decrease the effort required to analyze each patient. 
    \item Use feature importance charts, which help the physicians to understand which signals are the most important ones for the algorithms. This increases the validity of the trained models (explainability of results). 
\end{enumerate}

\begin{figure}[!htb]
    \centering
    \includegraphics[width=\columnwidth]{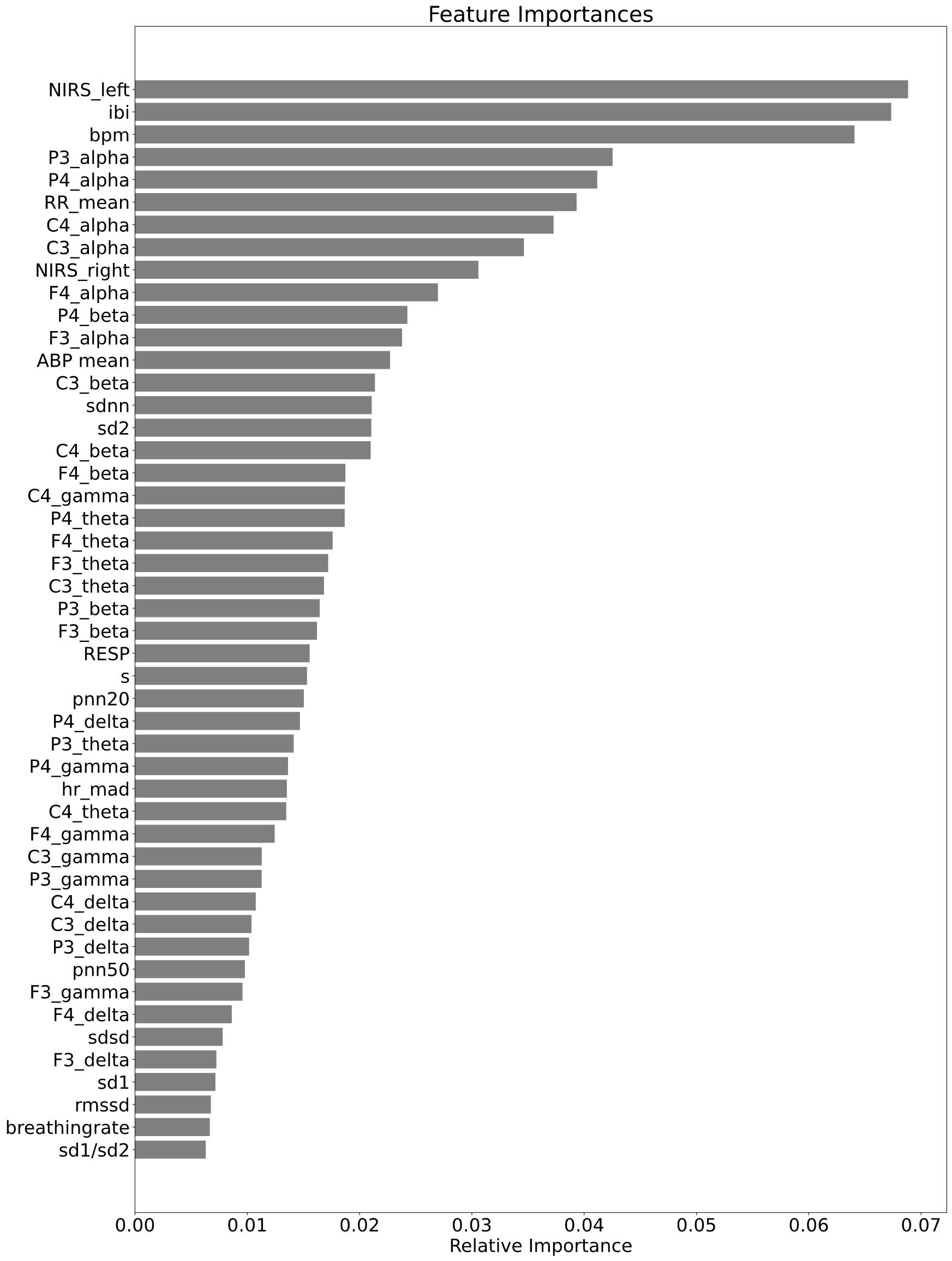}
    \caption{Feature importance chart for an example patient. The most important signals are NIRS (left side of the brain), Inter-beat interval (IBI), and Beats per Minute (BPM).}
    \label{fig:feature_importance}
\end{figure}

Figure \ref{fig:feature_importance} presents an example chart of feature importance for one patient. This kind of diagram provided the entire team with a good starting point for the discussion of the validity of results. 

After the analysis of the results, and discussing their empirical validity, the team was well equipped to discuss the confounding factors. 

\subsection{Mapping of confounding factors}
\noindent
In multidisciplinary studies, such as ours, identifying and reducing the confounding factors is a complex task. The confounding factors can originate from the medical domain and the software engineering domain respectively. They can only be identified when the team works together and when the team understands the entire data collection and analysis pipeline. 

The confounding factors in the medical domain can originate from the surgical procedures (e.g., using a stump blood pressure measurement instead of the arterial blood pressure, which causes disturbances in one of the signals), medication (e.g., anesthesia decreases the blood pressure) or labelling (e.g., the ability to identify important events in the OR). 

The confounding factors in the software engineering domain can originate from the data analysis procedures (e.g., machine learning algorithms used), feature extraction and data cleaning procedures (e.g., removing important data points when removing the noise ones), scaling up the software (e.g., introducing manual tasks) and the deployment (e.g., portability of libraries used). 

Our recommendations for finding the confounding factors efficiently are:
\begin{enumerate}
    \item Provide the summary charts that link the medical domain with the data analysis/software engineering domain. Using feature importance charts and t-SNE diagrams helps to get a summary view. This helps to start the discussion that is relevant for both domains.
    \item Provide detailed diagrams to enable deep-dive and traceability on a patient level. This helps the team to assess if certain data points are outliers or not (e.g., disturbances/artefacts in EEG signals). This increases the quality of the data and helps to find automated means for data quality assurance. 
    \item Go through the data from one patient for the entire surgical procedure. Such review of the data allows the team to understand what can go wrong in the procedure and change it. This helps to improve and optimize the procedures -- both in the OR and when analyzing the data.
    \item When finding confounding factors, discuss whether they can/should be mitigated by using analytical methods (e.g., data cleaning) or manual procedures (e.g., additional steps in the OR). 
\end{enumerate}

When identifying the confounding factors, the team can also identify potential new studies that can address these confounding factors. For example, in our case we identified the need to check how robust procedures are when used in more dynamic or emergent situation for example during an acute thrombectomy, where the patients need to be treated as soon as possible. Furthermore, we also found the need to design new studies with a group of patients not at risk for cerebral ischemia to establish the baseline for identifying events (to avoid the Hawthorne's effect). 

\subsection{Results from out studies on five patients}
\noindent
In order to illustrate the outcomes of such a collaboration, we present the final results of analysis of five patients. Each analysis was done using Random Forest and the goal was to recognize clinical events for each of these patients. Table \ref{tab:evaluation} presents the accuracy metrics for five patients included in our study.  

\begin{table}[!htb]
  \centering
  \begin{tabular}[width=\columnwidth]{||l|r|r|r||}
    \hline
    \textbf{Patient} & \textbf{Accuracy} & \textbf{Precision/ Specificity} & \textbf{Recall/ Sensitivity }\\ \hline
     Patient 1 & 0.98 & 0.98 & 0.98 \\ \hline 
     Patient 2 & 0.87 & 0.87 & 0.87 \\ \hline 
     Patient 3 & 0.89 & 0.92 & 0.92 \\ \hline 
     Patient 4 & 0.90 & 0.86 & 0.90 \\ \hline 
     Patient 5 & 0.82 & 0.94 & 0.94 \\ \hline 
  \end{tabular}
  \caption{Accuracy metrics for five evaluation patients.}
  \label{tab:evaluation}
\end{table}

The results vary from 0.82 to 0.98 in terms of accuracy. The ability to work together in this project led to defining an efficient data analysis pipeline. The data analysis pipeline take ca. 8-9 hours from the beginning of the surgery to the presentation of the results. The OR and post-op time is included (ca. 6-7 hours). 

\section{Recommendations}
\label{sec:recommendations}
\noindent
We found that effective and efficient collaboration between physicians and software engineers is essential for designing robust software systems. Involving software engineers as observers of the relevant clinical practice is crucial, just as involving physicians as observers of data analysis and software development. 

Therefore, we strongly recommend to involve both disciplines at the initiation of the study. This will lead to research questions that are relevant to both disciplines. In turn, that will lead to making contributions to both disciplines. 

We recommend to invest in learning about each other's disciplines in order to identify confounding factors, opportunities and to find best practices. By sharing recommendations about disseminating results, we can help to develop both disciplines in terms of scientific rigour and robustness of results. 

We also recommend to keep a journal of the project. Taking notes provides the team with the ability to reflect and retrospect on the progress of the project, which leads to recording new directions, ideas and analyses. 

\section{Conclusions}
\label{sec:conclusions}
\noindent
In this paper, we addressed the problem of how to establish an efficient collaboration between software engineers and physicians with the goal to design robust machine learning-based software systems for use in critical care. 

Our study resulted in finding 18 recommendations grouped into five phases -- starting from the initiation of the collaboration until the delivery of research results. One of the major findings was that investing in the design of the study, as well as joint execution of the pilot studies, are very important for the success of the design of software systems. We could also show that thanks to the collaboration, we were able to increase the accuracy of the machine learning models from 0.61 to 0.82, simply by identifying confounding factors, data cleaning based on clinical knowledge and statistics. 

Our further work include further development of our recommendations, including other contexts and other disciplines/domains. 

\section*{Acknowledgment}
\noindent
We would like to thank the patients for participating in our study. The study has been partially financed by CHAIR -- Chalmers AI Research Center and Wilhelm and Martina Lundgren science fund 2019-3078 as well as by the agreement between the Swedish Government and the county councils the ALF agreement ALFGBG-722182 and ALFGBG-936447. The study was performed in accordance with the most recent version of the Helsinki Declaration, and ethical consent for the study was obtained from the Swedish Ethical Review Authority,  Dnr 2020-00169.

\bibliographystyle{IEEEtran}
\bibliography{IEEEabrv,references}

\end{document}